# Quantum microwave photonic mixer with a large spurious-free dynamic range

Xinghua Li[1,2], Yifan Guo[1,2], Xiao Xiang[1,2], Runai Quan[1,2], Mingtao Cao[1,2], Ruifang Dong[1,2,3] *, Tao Liu[1,2,3], Ming Li[4,5,6] and Shougang Zhang[1,2,3]

[1]Key Laboratory of Time Reference and Applications, National Time Service Center, Chinese Academy of Sciences, Xi'an, 710600, China

[2]School of Astronomy and Space Science, University of Chinese Academy of Sciences, Beijing, 100049, China

[3]Hefei National Laboratory, Hefei 230088, People's Republic of China

[4]State Key Laboratory on Integrated Optoelectronics, Institute of Semiconductors, Chinese Academy of Sciences, Beijing, 100083, China

[5]School of Electronic, Electrical and Communication Engineering, University of Chinese Academy of Sciences, Beijing 100049, China

[6]Center of Materials Science and Optoelectronics Engineering, University of Chinese Academy of Sciences, Beijing 100190, China

## Abstract

As one of the most fundamental functionalities of microwave photonics, microwave frequency mixing plays an essential role in modern radars and wireless communication systems. However, the commonly utilized intensity modulation in the systems often leads to inadequate spurious-free dynamic range (SFDR) for many sought-after applications. Quantum microwave photonics technique offers a promising solution for improving SFDR in terms of higher-order harmonic distortion. In this paper, we demonstrate two types of quantum microwave photonic mixers based on the configuration of the intensity modulators: cascade-type and parallel-type. Leveraging the nonlocal RF signal encoding capability, both types of quantum microwave photonic mixers not only exhibit the advantage of dual-channel output but also present significant improvement in SFDR. Specifically, the parallel-type quantum microwave photonic mixer achieves a remarkable SFDR value of 113.6 dB·Hz$^{1/2}$, which is 30 dB better than that of the cascade-type quantum microwave photonic mixer. When compared to the classical microwave photonic mixer, this enhancement reaches a notable 53.6 dB at the expense of 8 dB conversion loss. These results highlight the superiority of quantum microwave photonic mixers in the fields of microwave and millimeter-wave systems. Further applying multi-photon frequency entangled sources as optical carriers, the dual-channel microwave frequency conversion capability endowed by the quantum microwave photonic mixer can be extended to enhance the performance of multiple-paths microwave mixing which is essential for radar net systems.

## Introduction

Microwave photonic mixers are widely adopted in various microwave and millimeter-wave systems, including radars [1, 2], wireless communication systems [3, 4], electronic warfare (EW) systems [5]. These mixers have been extensively developed to cater to specific mixing functions required for different applications. For instance, the microwave photonics-based image rejection mixer (MP-IRM) has been designed to eliminate undesired signals from the image, offering a larger bandwidth compared to traditional electrical IRMs [6, 7]. Utilizing a dual-polarization quadrature phase shift-keying modulator (DP-QPSKM), the microwave photonic mixer with suppressed spurs and dispersion immunity has been reported, which is favorable for radio-over-fiber (RoF) communication systems [8]. For hybrid macro-micro cellular systems, a reconfigurable microwave photonic mixer has been developed, providing versatile functions such as single-ended dispersion immune mixing, I/Q frequency down-conversion, image rejection mixing, and double-balanced mixing [9]. In recent years, integrated microwave photonic mixers with high-performance have also been developed [10-12].

For a microwave photonic mixer, the spurious-free dynamic range (SFDR) is a crucial parameter that defines the range between the highest and lowest power levels at which a microwave frequency mixer can operate effectively [13]. This parameter has long been recognized as one of the most challenging aspects in the field of microwave photonics. Two primary strategies commonly used to enhance the SFDR involve the utilization of low-noise laser diodes (LDs) [14] and amplifiers [15] to reduce the system noise floor, and the integration of high-performance intensity modulator to mitigate the inherent modulation nonlinearity. For example, by utilizing a highly linear optical modulator, a microwave frequency downconverter with a SFDR over 120 $dB \cdot Hz^{2/3}$ in terms of cross modulation distortion was reported [16]. By developing a quantum-well modulator with low half-wave voltage and high linearity, Li et al. demonstrated 28 dB improvement in distortion levels over that using a $LiNbO_3$ Mach-Zehnder (MZ) modulator [17]. To further combat the inadequate SFDR inherent in the optical intensity modulation, a phase modulated RF/photonic link was demonstrated, which achieved a record-breaking SFDR of ~129.3 $dB \cdot Hz^{2/3}$ [18]. However, the highly linear phase modulation commonly involves the utilization of dual-parallel modulators alongside sophisticated optical phase-locked loop and balanced detection techniques [18, 19]. It also has a very restricted bandwidth due the optical filter bandwidth [20, 21]. Additionally, its inability to access ultra-weak microwave photonic frequency mixing becomes another crucial bottleneck [22].

To address this issue, the single-photon microwave photonics (SP-MWP) technique was reported, which leverages super low-jitter and high-sensitivity single-photon detectors (SPDs) to enable single-photon-carried RF signal detection and processing with a bandwidth determined by the timing jitter of the SPDs [23]. Further combining energy-time entangled biphoton source [24, 25] as the optical carrier, quantum microwave photonics (QMWP) was explored, demonstrating remarkable capabilities in nonlocal RF signal modulation, phase shifting, and multi-tap transversal frequency filtering functions with excellent resistance to dispersion-associated frequency fading in radio-over-fiber (ROF) applications [26, 27, 28]. In contrast to classical microwave photonics (MWP), QMWP technology eliminates the significant frequency fading caused by dispersion associated with the carrier bandwidth. Consequently, it overcomes the limitations to the carrier bandwidth in classical MWP links. This paper further demonstrates its proficiency in microwave signal mixing and its potential superiority.

In this paper, two types of QMWP mixers based on the configuration of the intensity modulators are presented: cascade-type and parallel-type. To assess the linearity performance of the output mixed microwave signal, which is key for many sought-after microwave photonic applications such as radars, the spurious free dynamic range (SFDR) is utilized. Compared to the classical MWP mixer, both types of QMWP mixers show significantly improved SFDR in terms of the second harmonic distortion by more than 20 dB, despite at a modest cost of an 8 dB conversion loss. Particularly, the parallel-type QMWP mixer demonstrates an ultra-high SFDR of approximately 113.6 dB·Hz$^{1/2}$, which is even improved by 30 dB in comparison with the cascade-type QMWP mixer. Furthermore, with the commercial deployment of fifth-generation (5G) communication networks and the application of multi-input multi-output (MIMO) systems [30], a multi-channel mixer is urgently needed to realize the reception and processing of multi-channel signals. Thanks to the nonlocal RF signal modulation capability, both types of QMWP mixers provides the ability of dual-channel signal reception and processing. When a multi-photon frequency entangled source can be generated and applied as the optical carrier, the parallel frequency conversion between multiple base stations and a central station in a radar net system holds immense potential for advancement.

## Theoretical principle

In principle, the frequency mixing can be regarded as a nonlinear process, by which new frequency components are created. In our experiment, the energy-time entangled biphotons [31, 32] act as the optical carrier, the corresponding joint temporal wave function can be given by [27]

$$\Psi(t_1, t_2) \propto \exp[-i(\omega_{s,0} t_1 + \omega_{i,0} t_2)] \exp\left[-\frac{\sigma^2(t_1 - t_2)^2}{2}\right]. \tag{1}$$

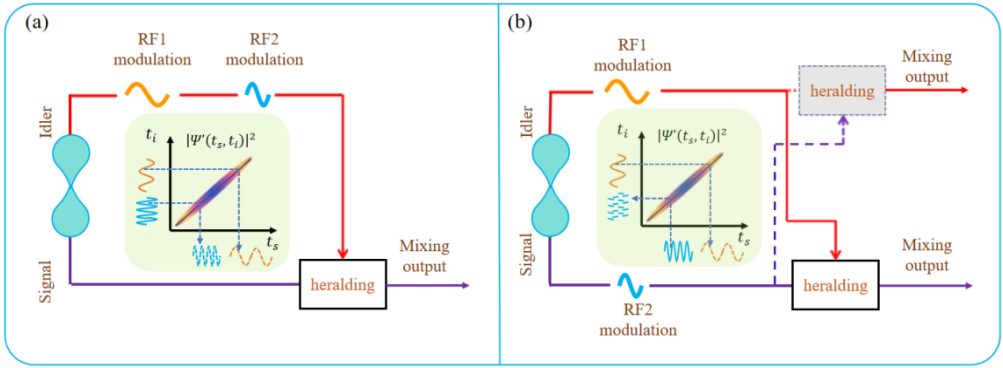

Fig. 1 Scheme of QMWP mixers with the RF modulators in the configuration of (a) cascade-type; (b) parallel-type.

Where $t_1$ and $t_2$ denote the temporal coordinates of the signal and idler photons, respectively. $\omega_{s(i),0}$ denotes the center angular frequency of the signal (idler) photons, while $\sigma$ represents their single-photon spectral widths.

For the cascade-type QMWP mixing case, as shown in Fig. 1(a), two RF signals with frequencies $\omega_{RF1}$ and $\omega_{RF2}$ are sequentially intensity modulated onto the idler photons. The transfer function is given by

$$M(t_2) = [1 + A_1 \cos(\omega_{RF1} t_2)] \times [1 + A_2 \cos(\omega_{RF2} t_2)]. \tag{2}$$

Where $0 < A_{1(2)} \leq 1$ represents the modulation magnitude. After the modulation, the two-photon wave function transforms to

$$\Psi'(t_1, t_2) = M(t_2)\Psi(t_1, t_2). \tag{3}$$

The square modulo of the two-photon wave function defines the second-order Glauber correlation function ($G^{(2)}$) [32], i.e., $G^{(2)} \equiv |\Psi'(t_1, t_2)|^2$, which is measured by the biphoton coincidence

distribution. The expression for $|\Psi'(t_1, t_2)|^2$ is given by:

$$|\Psi'(t_1, t_2)|^2 = \exp[-\sigma^2(t_1 - t_2)^2]\left(1 + 2A_1\cos(\omega_{RF1}t_2) + \frac{A_1^2}{2}(1 + \cos(2\omega_{RF1}t_2))\right)$$

$$\times \left(1 + 2A_2\cos(\omega_{RF2}t_2) + \frac{A_2^2}{2}(1 + \cos(2\omega_{RF2}t_2))\right). \quad (4)$$

By performing marginal integration over the temporal coordinates of $t_2$, the temporal wave function of the heralded signal photons can be expressed as:

$$\varphi'_s(t_1) \sim \int dt_2 |\Psi'(t_1, t_2)|^2 = \frac{\sqrt{\pi}}{\sigma}\left[1 + \alpha e^{-\frac{\omega_1^2}{4\sigma^2}}\cos[\omega_{RF1}t_1] + \beta e^{-\frac{\omega_2^2}{4\sigma^2}}\cos[\omega_{RF2}t_1] + \right.$$

$$\left. \gamma e^{\frac{-(\omega_1-\omega_2)^2}{4\sigma^2}}\cos[(\omega_{RF1} - \omega_{RF2})t_1] + \gamma e^{\frac{-(\omega_1+\omega_2)^2}{4\sigma^2}}\cos[(\omega_{RF1} + \omega_{RF2})t_1] + h(t_1)\right] \quad (5)$$

Where $\alpha, \beta$ and $\gamma$ are expressed by $\frac{A_1(2+A_2^2)}{2}, \frac{A_2(2+A_1^2)}{2}$ and $A_1 A_2$, respectively. In Eq. (5), two frequency mixing components, the sum frequency component $(\omega_{RF1} + \omega_{RF2})$ and the difference frequency component $(\omega_{RF1} - \omega_{RF2})$ are generated. The term $h(t_1)$ represents the unwanted spurious frequency components, which are related to the high-order harmonics term (its expression is given in the Appendix). Similarly, the temporal waveforms of the idler photons derived after integration over $t_1$ can be expressed as:

$$\varphi'_s(t_2) \sim \int dt_1 |\Psi'(t_1, t_2)|^2 = \frac{\sqrt{\pi}}{\sigma}[1 + 2\alpha\cos(\omega_{RF1}t_2) + 2\beta\cos(\omega_{RF2}t_2) + 2\gamma\cos[(\omega_{RF1} - \omega_{RF2})t_2] + 2\gamma\cos[(\omega_{RF1} + \omega_{RF2})t_2] + h(t_2)], \quad (6)$$

where $h(t_2)$ represents the unwanted spurious frequency components in the idler photon waveform (also given in the Appendix). Comparing Eq. (5) and Eq. (6), it can be observed that the RF modulations on the idler photons can be nonlocally mapped onto their entangled counterparts, the signal photons. Thus, the frequency mixing function can occur in the non-modulated path.

For the parallel-type QMWP mixing case, as shown in Fig. 1(b), two RF signals with frequencies $\omega_{RF1}$ and $\omega_{RF2}$ are intensity modulated onto the idler photons and signal photons in parallel type, respectively. The transfer function for this system is given by:

$$M(t_1, t_2) = [1 + A_1\cos(\omega_{RF1}t_1)] \times [1 + A_2\cos(\omega_{RF2}t_2)]. \quad (7)$$

After the modulation, the two-photon wave function transforms to:

$$\phi'(t_1, t_2) = M(t_1, t_2)\Psi(t_1, t_2). \quad (8)$$

The square modulo of the two-photon wave function is expressed by:

$$|\phi'(t_1, t_2)|^2 = \exp[-\sigma^2(t_1 - t_2)^2]\left(1 + 2A_2\cos(\omega_{RF2}t_2) + \frac{A_2^2}{2}(1 + \cos(2\omega_{RF2}t_2))\right)$$

$$\left(1 + 2A_1\cos(\omega_{RF1}t_1) + \frac{A_1^2}{2}(1 + \cos(2\omega_{RF1}t_1))\right). \quad (9)$$

Similarly to the derivation process of the cascade-type RF modulation, the temporal waveforms of the idler photons and signal photons are expressed as:

$$\rho_s(t_1) \sim \int dt_2 |\phi'(t_1, t_2)|^2 = \frac{\sqrt{\pi}}{\sigma}\left[I + \alpha\cos(\omega_{RF1}t_1) + \beta e^{-\frac{\omega_{RF2}^2}{4\sigma^2}}\cos[\omega_{RF2}t_1] + 2\gamma e^{-\frac{\omega_{RF2}^2}{4\sigma^2}}\cos[(\omega_{RF2} - \right.$$

$$\left. \omega_{RF1})t_1] + 2\gamma e^{-\frac{\omega_{RF2}^2}{4\sigma^2}}\cos[(\omega_{RF2} + \omega_{RF1})t_1] + h'(t_1)\right]. \quad (10)$$

$$\rho_i(t_2) \sim \int dt_1 |\phi'(t_1,t_2)|^2 = \frac{\sqrt{\pi}}{\sigma}\left[I + \beta\cos(\omega_{RF2}t_2) + \alpha e^{-\frac{\omega_{RF1}^2}{4\sigma^2}}\cos[\omega_{RF1}t_2] + 2\gamma e^{-\frac{\omega_{RF1}^2}{4\sigma^2}}\cos[(\omega_{RF2} - \omega_{RF1})t_2] + 2\gamma e^{-\frac{\omega_{RF1}^2}{4\sigma^2}}\cos[(\omega_{RF2} + \omega_{RF1})t_2] + h'(t_2)\right]. \tag{11}$$

Where $I = \left(1 + \frac{A_2^2}{2}\right)\left(1 + \frac{A_1^2}{2}\right)$, the term $h'(t_{1(2)})$ also represents the spurious frequency components, which are related to the high-order harmonics term (given in the Appendix). From Eq. (11), it can be observed that the frequency mixing is achieved when the spatially separated RF$_1$ and RF$_2$ signals are modulated onto different photons (signal photons and idler photons). The frequency entanglement between the paired photons enables the spatially separated RF$_1$ and RF$_2$ signals to be nonlocally mapped onto their counterpart optical carriers and then dual-channel frequency mixing from the parallel-type QMWP mixer is output.

**Experimental Setup**

In Fig. 2 (a), energy-time entangled biphotons (signal and idler photons) are generated from a 10 mm long type-II PPLN waveguide [24]. These photon pairs are produced by pumping the waveguide with a continuous-wave laser at a wavelength of 780 nm. The generated signal and idler photons act as optical carriers and undergo RF signal modulation, as detailed in Fig. 2 (b) and 2 (c). After modulation, the entangled biphotons are detected by low-jitter superconductive nanowire single-photon detectors (SNSPD1 & SNSPD2, Photec), which have a timing jitter of approximately 50 ps. The outputs of the two SNSPDs are then fed into a Time-Correlated Single Photon Counting (TCSPC) module (PicoQuant Hydraharp 400). This TCSPC module operates in the Time Tagged Time-Resolved (TTTR) T3 mode with the time-bin resolution set at 8 ps. A TTL signal with frequency of 20 MHz, generated from the arbitrary waveform generator (AFG3252, Tektronix), is used as the trigger signal for the TCSPC.

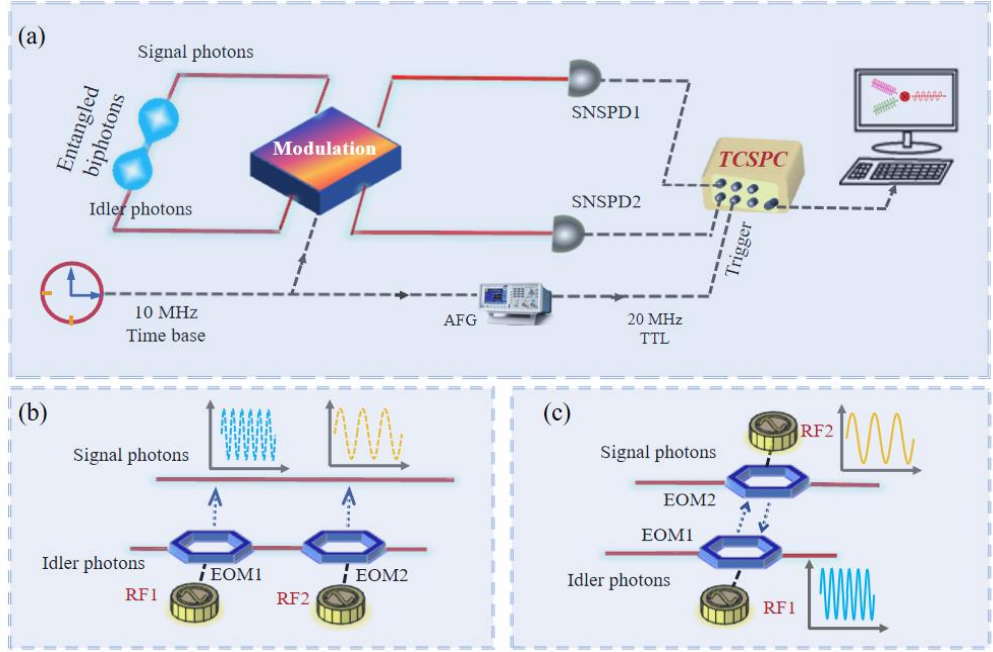

Fig. 2 Experimental diagram of a QMWP mixer system. EOM1 and EOM2: electro-optic modulator for intensity modulation; AWG: arbitrary waveform generator; SNSPD1 and SNSPD2, superconducting nanowire single-photon detectors; and TCSPC, time-correlated single-photon counter.

In the case of a cascade-type QMWP mixer, as shown in Fig. 2 (b), the idler photons undergo a sequential intensity modulation through two Mach-Zehnder modulators (MZM, PowerBit™ F10-0 from Oclaro) named EOM1 and EOM2. These modulators are respectively driven by high-speed sinusoidal RF signals from two signal generators (DSG3000B, RIGOL Technologies). To establish phase stabilization between the RF signal generator and the TTL signal, a 10 MHz time base serves as an external reference. On the other hand, in the case of a parallel-type QMWP mixer, as shown in Fig. 2 (c), the idler photons and signal photons are intensity modulated by EOM1 and EOM2, respectively.

**Results and Discussion**

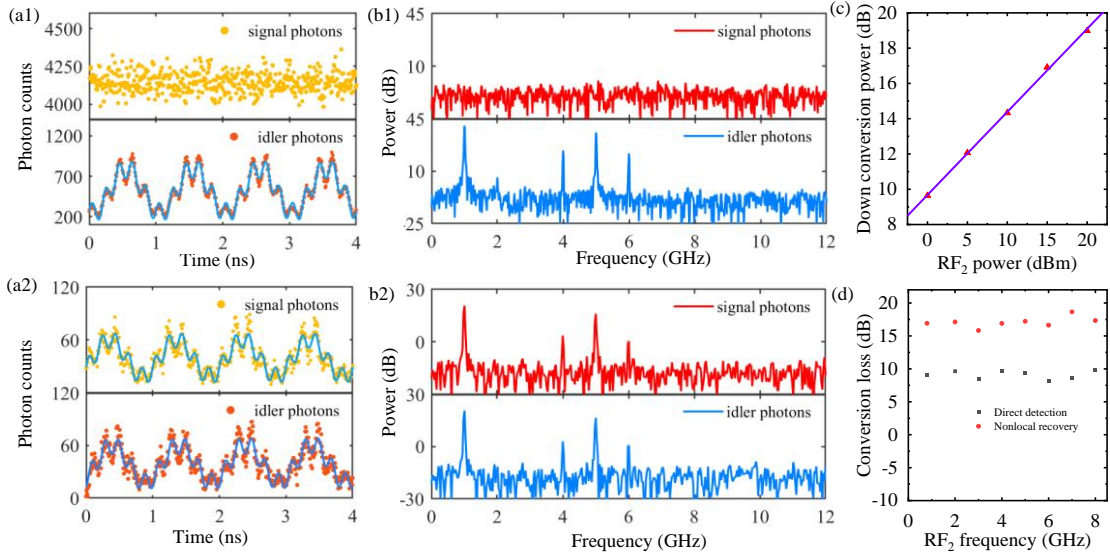

Fig. 3 The waveforms of the signal and idler photons over time, and their corresponding power spectra, based on the cascade-type QMWP mixer through direct detection: (a1) - (b1) coincidence-based post-selection: (a2) - (b2). (c) The dependence of cascade-type QMWP mixer output power on the input $RF_2$ power. (d) The measured conversion loss for classical cascade mixer (black squares) and cascade-type QMWP mixer (red dots) as a function of the $RF_2$ frequency.

The nonlocal mapping of the RF modulation on the idler photons to the signal photons is firstly investigated for the cascade-type QMWP mixer by setting the $RF_1$ signal at 1 GHz with a modulation power of 8 dBm and the $RF_2$ signal at 5 GHz with a modulation power of 10 dBm. Fig. 3 (a1) depicts direct measurements of the signal and idler photon waveforms over time. The solid curve represents the theoretical fitting based on Eq. (2), which aligns well with the experimental results. After applying the discrete Fourier transform (DFT) to the waveforms, the corresponding power spectra of the signal and idler photons are shown in Fig. 3 (b1). The idler spectrum displays two RF fundamental frequencies, as well as the expected sum and difference frequencies. This observed mixing in the idler photons is consistent with classical cascade mixing process [33]. However, the power spectrum of the signal photons is free from any frequency components. It is well-known that frequency-correlated photon waveforms can be selectively measured using time-correlation coincidence measurement [34]. Thus, by applying post-selection to the biphoton coincidence distribution, the RF modulation and frequency mixing phenomenon can also be nonlocally observed in the signal photon waveforms. This indicates that any time/frequency-related manipulation applied to the idler photons will be nonlocally mapped onto their relevant signal photons [35, 36]. To showcase the nonlocal mapping, Fig. 3 (a2) displays the recovered waveforms of the signal and idler photons after applying coincidence-based post-selection, which matches well with their theoretical fitting based on Eq. (5) and Eq. (6). Their corresponding power spectra are also illustrated in Fig. 3 (b2). As expected, the power spectrum of the signal photons is similar to that of the idler photons. The output power of the difference frequency component, which is nonlocally measured from the signal photons, as a function of the input $RF_2$ signal power is shown in Fig. 3 (c). It demonstrates a perfect linear relationship with the increase in input $RF_2$ power. By fixing the RF powers and varying the frequency of $RF_2$, the conversion loss of the frequency down conversion is measured and shown in Fig. 3 (d). It can be observed that the conversion loss of the nonlocally achieved difference frequency signal is approximately 18 dB, which is 8 dB more than that of directly achieved difference frequency signal. This additional conversion loss is primarily due to the post-selection process, which reduces the number of carrier photons in 180 s from $3.2 \times 10^6$ under direct detection to $1.44 \times 10^5$, leading to a loss of 13.4 dB. It's worth noting that, the conversion loss is not linearly related to the optical power of carrier photon [13]. Therefore, there is no direct correspondence between the additional conversion loss and the photon loss induced by the post-selection.

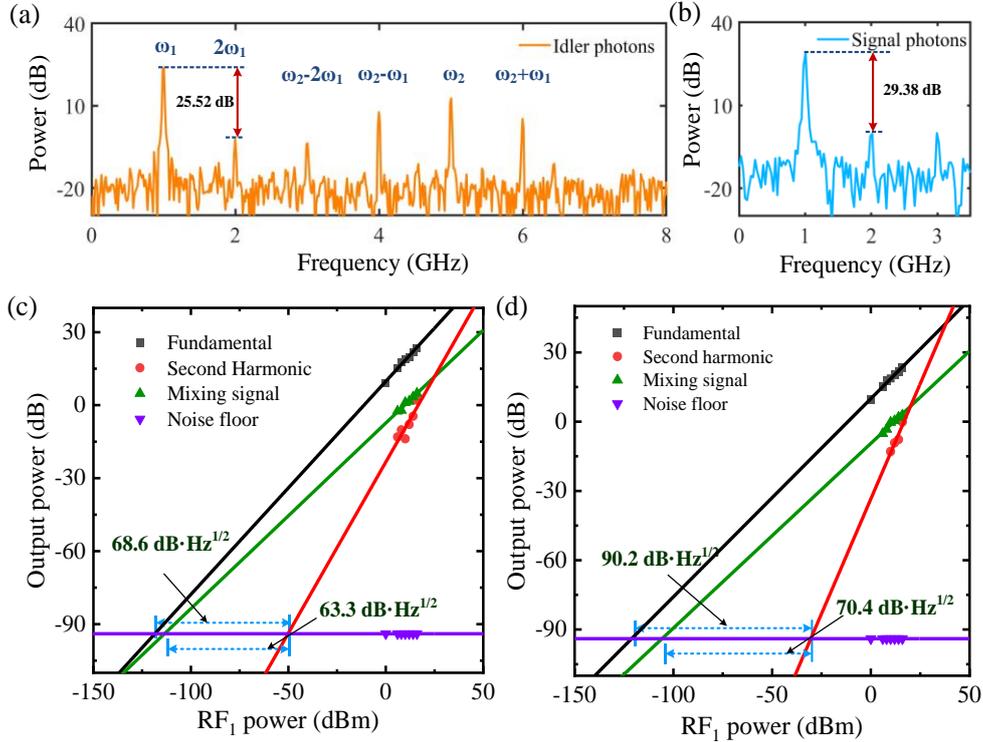

Fig. 4 Power spectra of the idler photons and signal photons with $RF_1$ and $RF_2$ signals being sequentially modulated onto the idler photons at frequencies of 1 GHz and 5 GHz respectively, whose input powers are set at 14 dBm and 10 dBm. (a) Direct detection. (b) Nonlocal recovery. The measured output powers of the fundamental signal ($\omega_1$), frequency mixing signal ($\omega_2 - \omega_1$), and second harmonic signal ($2\omega_1$) as a function of the input power of $RF_1$ signal

for the cases of (c) under direct detection and (d) after applying coincidence-based post-selection.

With the input $RF_1$ signal power increased to 14 dBm, while retaining the $RF_2$ signal power at 10 dBm, Fig. 4(a) illustrates the RF signal power spectrum carried by the idler photons under direct detection. It shows the emergence of various unwanted frequency components alongside the desired frequency mixing signals $\omega_2-\omega_1$ and $\omega_2+\omega_1$, including the second harmonic signal $2\omega_1$, its related frequency mixing signals $\omega_2-2\omega_1$, and $2\omega_2-\omega_1$. Note that, the utilized SNSPD has a timing jitter of approximately 50 ps (FWHM), which sets the maximum detection frequency bandwidth to be about 8.8 GHz [23] according to the Fourier transformation law. As a result, the second harmonic term of $RF_2$ and its related term can be filtered out by the detection process. Thus, the second harmonic of $RF_1$ signal becomes the primary factor that limits the dynamic range of the microwave mixer. To improve the performance of the mixer, it is necessary to suppress this second harmonic of the $RF_1$ signal to ensure a wider dynamic range and better overall functionality. Under direct detection, the power ratio of the fundamental signal $\omega_1$ to the second harmonic signal $2\omega_1$ is measured to be 25.52 dB. After applying the coincidence based post-selection technique (as shown in Fig. 3(b)), this power ratio increases to approximately 29.38 dB, indicating a suppression of spurious signals by nearly 4 dB.

With regard to the two cases of under direct detection and under post-selection, the powers of the fundamental signal ($\omega_1$), frequency mixing signal ($\omega_2 - \omega_1$), and second harmonic signal ($2\omega_1$) as a function of the input $RF_1$ power are respectively measured and illustrated in Fig. 4(c) and (d). After normalizing the noise power to 1 Hz, the noise floor and the SFDRs of these three frequency signals can be extrapolated by applying linear fitting to these measurements. It can be observed that, the second-order SFDR of the fundamental signal is increased from 68.9 dB·Hz$^{1/2}$ under direct detection to 90.2 dB·Hz$^{1/2}$ under post-selection, and the second-order SFDR of the difference frequency signal is increased from 63.3 dB·Hz$^{1/2}$ under direct detection to 70.4 dB·Hz$^{1/2}$ under post-selection.

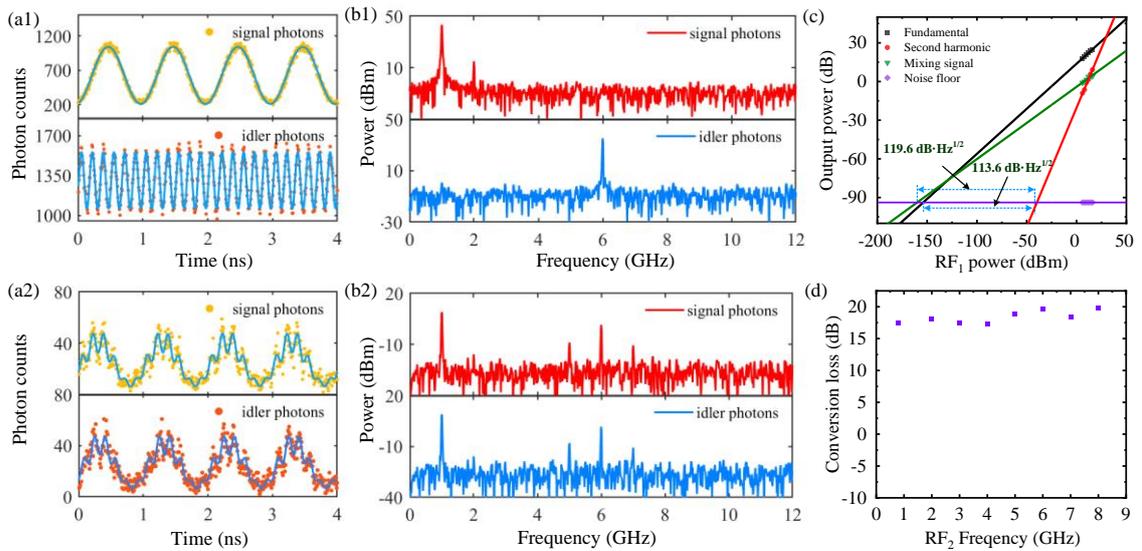

Fig.5 The waveforms of the signal and idler photons over time, and their corresponding power spectra, based on the parallel-type QMWP mixer through direct detection: (a1) - (b1) coincidence-based post-selection: (a2) - (b2). (c)

The measured output powers of the fundamental signal ($\omega_1$), frequency mixing signal ($\omega_2 - \omega_1$), and second harmonic signal ($2\omega_1$) as a function of the input power of RF$_1$ signal. (d) The measured conversion loss as a function of the RF$_2$ frequency.

Subsequently, we explore the performance of the parallel-type QMWP mixer as shown in Fig. 2 (c). The RF$_1$ signal loaded on EOM$_1$ is set to be 1 GHz with a modulation power of 8 dBm, and the RF$_2$ signal loaded on EOM$_2$ is set to be 6 GHz with a modulation power of 10 dBm. Similarly to the cascade-type QMWP mixer case, Fig. 5 (a1) depicts direct measurements of the signal and idler photon waveforms over time, the solid lines are the theoretical fitting curve. After applying discrete Fourier transform (DFT) to the waveforms, the corresponding power spectra of the signal and idler photons are shown in Fig. 5 (b1). It can be observed that both spectra display only their individually carried fundamental RF frequencies (1 GHz and 6 GHz). Fig. 5 (a2) displays the recovered waveforms of the signal and idler photons after applying coincidence-based post-selection, which match well with their theoretical fittings based on Eqs. (10) and (11). Their corresponding power spectra are also illustrated in Fig. 5 (b2). It is readily seen that, the nonlocal mapping of the RF modulation between the entangled photon pairs not only happens on the fundamental components but also on the frequency mixing components. Fig. 5 (c) shows the measured powers of the fundamental signal ($\omega_1$), the difference frequency signal ($\omega_2 - \omega_1$), and the second harmonic signal ($2\omega_1$) as a function of the input RF$_1$ power. The SFDRs are extrapolated by applying linear fitting to the measured data. The results show that second-order SFDR of the fundamental signal and the down-converted signal are significantly improved to 119.6 dB·Hz$^{1/2}$ and 113.6 dB·Hz$^{1/2}$, respectively. Fig. 5 (d) displays the measured conversion loss of the parallel-type QMWP mixer system at different modulation frequencies of RF$_2$, which is approximately 18 dB and consistent with that of the cascade-type QMWP mixer. Compared with the second-order SFDR of the classical cascade mixer in Fig. 4 (c), an improvement of 53.6 dB is achieved at the cost of an 8 dB conversion loss. Moreover, the nonlocal parallel mixing technique can be effectively expanded to multi-channel parallel modulations by employing multi-photon frequency entangled sources as optical carriers. With the help of this nonlocal RF mapping technique, it facilitates a parallel processing function between each photon carrier channel, enabling the reception and processing of multi-channel signals in MIMO systems [37, 38]. This technique holds great promise for applications in fifth-generation (5G) communication networks and radar networks.

## Conclusion

In conclusion, this paper introduces two novel QMWP mixers that leverage nonlocal RF signal encoding, benefiting from energy-time entangled biphoton sources as the optical carriers. Both mixers demonstrate excellent linearity between the output and input RF powers, with the parallel-type QMWP mixer showcasing superior performance in SFDR. The parallel-type QMWP mixer achieved an ultra-high second-order SFDR of approximately 113.6 dB·Hz$^{1/2}$, surpassing that of the cascade-type QMWP mixer by 30 dB at the same conversion loss level. When compared to the classical cascade-type microwave photonic mixing technique, the parallel-type QMWP mixer achieves a remarkable enhancement of 53.6 dB in the second-order SFDR, while incurring an additional conversion loss of only 8 dB. Furthermore, the nonlocal QMWP mixing capability demonstrated in this study can be extended to multiple EOMs modulation by utilizing multi-photon frequency entangled sources as optical carriers. This opens up possibilities for parallel frequency conversion between multiple base stations and a central station. These findings highlight the potential of quantum microwave photonic mixers in improving the performance and efficiency of microwave communication systems.

## Acknowledgement

The National Natural Science Foundation of China (Grant Nos. Grant Nos. 12033007, 61801458, 12103058, 12203058, 12074309, 61875205), the Key Project of Frontier Science Research of Chinese Academy of Sciences (Grant No. QYZDB-SSW-SLH007), the Strategic Priority Research Program of CAS (Grant No. XDC07020200), the Youth Innovation Promotion Association, CAS (Grant No. 2021408,2022413).